\documentclass[12pt]{article}
\usepackage{hyperref,xcolor}
\usepackage{cite}
\usepackage{amsmath}
\usepackage{graphicx, tikz}
\usepackage{amssymb}
\usepackage{hyperref}
\usepackage{footnote}
\makesavenoteenv{minipage}

\newcommand{\mc}{\mathcal}
\newcommand{\p}{\partial  }

\renewcommand{\title}[1]{%
    \bigskip%
    \begin{center}%
    \Large\bf #1%
    \end{center}%
    \vskip .2in}

\renewcommand{\author}[1]{%
    {\begin{center}
    #1
    \end{center}}}
\newcommand{\address}[1]{\vspace{-1.7em}\vspace{0pt}
    {\begin{center}
    \it #1
    \end{center}}}

\begin{document}

\begin{titlepage}
%\title{Ostrogradski Ghosts and $\mathcal{PT}$ Symmetry theories}
%\title{$\mathcal{PT}$ symmetric treatment to Ostrogradski instability }
%\title{Cuscuton gravity with Noether symmetry and dynamical analysis  }
\title{Cyclic cosmology from Cuscuton-Gallileons obeying Lie   point  transformations }
\author
{
Biswajit Paul $\,^{\rm a,b}$,
Pushpendra Kumar Singh  $\,^{\rm a,c}$,

}
 %\footnote{Also, Visiting Associate, S. N. Bose National Centre for Basic Sciences, JD Block, Sector III, Salt Lake City, Kolkata -700 098, India  } 
%,
\address{$^{\rm a}$ National Institute of Technology Agartala \\
 Jirania, Tripura -799 055, India }

%\address{$^{\rm c}$Department of Physics, Barasat Government College,\\Barasat, West Bengal }
\footnote{
%\address 
{$^{\rm c}$\tt pushpendra.ag.mp@gmail.com} ,
{$^{\rm b}$\tt biswajit.thep@gmail.com,}
 {$^{\rm b}$\tt biswajitpaul.phy@faculty.nita.ac.in}}
\begin{abstract}
Spacetime transformations in any physically viable theory should follow  Lie Point symmetry.  In this work, we explore the Cuscuton model extended to Galileons, as introduced by de Rham et al in \cite{Rham2017}. We find the true degrees of freedom by converting the model into an equivalent first order model. Despite being a higher derivative model, it possesses only \textit{two} degrees of freedom. We calculate the Noether symmetry parameters corresponding to Lie point transformations, which lead to the vanishing of the original Cuscuton term's coefficient and restrict the potential to an exponential form. Interestingly, the coefficient corresponding to the original Cuscuton term vanish. Additionally, we also use the  Killing analysis to find out the charges corresponding to the Killing vectors and the Killing tensors. The cosmological implications are examined through dynamical analysis, revealing that under the condition where the coefficient $a_2$ vanishes, the equation of state parameters exhibit damped oscillatory behavior .

\end{abstract}

\end{titlepage} 
%\newpage
\section{Introduction}\label{intro}
%\textbf{Requirements for DE: }

%textbf{Requirements for Cuscuton:}
The cuscuton model was introduced in \cite{Afshordi2007} as a noncanonical term added to usual Einstein-Hilbert action. It was shown to be  an incompressible K-essence fluid model and to remain causal despite having infinite sound speed. Soon after the introduction, these models are used in many fields like in braneworld scenario \cite{Andrade2019, Mylova2024, Bazeia2023}, dark energy \cite{Bazeia2025, Iyonaga2020}, Galileon cosmology \cite{Maeda2022}, inflation \cite{Channuie2023, Bartolo2022}, solitons \cite{Lima2023}.The dynamical stability of cuscuton-like Lagrangians were also considered in \cite{Bessada2012}.  In \cite{Dehghani2025}, it is shown from analytical as well as numerical calculations that when Cuscutons are considered in non-singular bouncing cosmology the results are consistent in infrared as well as UV region. Hamiltonian analysis of various Cuscuta models were considered in \cite{Mansoori2023}. It is also shown that, the presence of the Cuscuton term in the action lead to change in braneworld configurations  by effectively increasing the height of the stability barrier\cite{Rosa2022}. 

Any modification to the well known gravity theories often lead to increase in degrees of freedom, ac common feature for many scalar-tensor theories.    Based on this, one might expect models containing a Cuscuton term to exhibit additional degrees of freedom. Interestingly, the Cuscuta theories, from the beginning, were constructed so that the field  becomes non-dynamical and as a result  no extra degrees of freedom are added\cite{Afshordi2007}.  Hamiltonian analysis in ADM formalism was performed for theories of gravity with Cuscuton term and confirmed to possess  \textit{two} degrees of freedom, as in the usual case \cite{Gomes2017, Mukohyama2019}.  The Cuscuta theories were extended for general Horndeski theories and Hamiltonian analysis was performed in \cite{Iyonaga2018} to find out the degrees of freedom. Interestingly, here also, the degrees of freedom count in the unitary gauge were found to be {two}.  

It is well known that most of the scalar-tensor theries or modified $f(R)$ theories are higher derivative(HD) in nature. By this, we mean that the equations of motion contain time derivatives of the fields more than two . Higher derivative theories are are of significant interest and appear in various branches of physics including general relativity (GR) \cite{Otinger2020, Sotiriou2010, Hindawi1996}, Cosmology \cite{Carloni2019, Goswami2008}, String theory \cite{Peeters2002, Becker2010, Howe2003} etc.  True  HD theories typically introduce extra degrees of  freedom called Ostrogradski ghost fields and these malign the corresponding quantum picture of the theory. Therefore, it is crucial to examine the degrees of freedom count of Cuscuton models from the perspective of higher derivative terms, especially if such terms are present, to ensure the absence of these pathological modes.

 Another important aspect of counting of degrees of freedom  comes from the symmetry  considerations related to the Noether's theorem. One generally proceeds to check the invariability of either the lagrangian or the associated Euler-Lagrange equations of motion which are generally non-linear partial differential equations. The goal is to find out the corresponding symmetry group under which the action is invariant. To achieve this, infinitesimal transformation parameters are introduced, and the first prolongation of the Lagrangian is computed, leading to a conservation equation. These give a set of partial differential equations solving which we can find   of the symmetry parameters \cite{BlumanB}.   
  Usually physically meaningful theories in General Relativity obeys contact symmetries which preserve the structure of the manifold and equations of motion.  These symmetries are  one of the Lie symmetries  and that requires finding out the parameters which are infinitesimal in nature. Since the models of  interest here may involve even higher orders of derivative, it is instructive to consider rather Lie-Backlund symmetries. Lie symmetry parameters are calculated in various models relating to general relativity \cite{Christodoulakis2014, Capozziello1997 }, $f(R)$-gravity \cite{Paliathanasis2016, Paliathanasis2016a,  Vakili2008}, teleparallel $f(T)$ gravity \cite{Wei2012}, cosmology\cite{Dialektopoulos2019, Dimakis2017, Paliathanasis2016}, Gauss-Bonnet gravity \cite{Sanyal2011}  etc. Therefore it will be  important as well as interesting aspect to find the Lie-Backlund symmetry parameters for the infinitesimal transformations of the Cuscuton gravity models. There is another way to check the preservation of the metric structure of the spacetime requires another Lie symmetry viz the Killing   symmetry. Killing vectors are associated to the specific form of the metric which in turn may yield corresponding conserved charges. Not only that, one can go beyond and calculate the Killing tensors from the generalized Killing equation. The conserved quantities can be constructed accordingly.  
 
%\textbf{The Noether symmetry and connection to compactness of the manifold:}
%Requirements of the oscillatory model

From standard cosmology, it is inherently required that the Universe has a beginning called the Big Bang and that consequently   leads to the appearance of singularity.  The Big Bang model though a very successful model for explaining the current observed acceleration but it cannot address several fundamental  issues like  the flatness problem,  homogeneity problem, the monopole problem  etc.  The singularity problem exists both in GR and cosmology as the initial conditions till remain unknown.   These challenges motivated to the introduction of alternative models of the Universe and among them one of the very successful models is the Oscillatory model. The Oscillatory models  were there from the foundational discussions on various models of cosmology as proposed by Tolman \cite{tolman1931}, Lamitre \cite{lamaitre}.  Accordingly, another cyclic model of the Universe, considered to be an alternative to Tolman's model \cite{tolman1931}, is recently  proposed by Steinhardt et al \cite{Steinhardt2002, Steinhardt_2002a}. This is one, in which the Universe undergoes periodic 'bangs' and `crunches',  has colliding branes in a negative potential of scalar fields. Consequently, it was shown that these oscillatory models can solve the coincidence problem \cite{Steinhardt2002, Chang2012, Doelson_2000} .  Since then oscillatory models  have been applied in $\Lambda CDM$ models \cite{Pan}, modified gravity \cite{Pavlovic_2017} , $f(R,T)$ gravity \cite{Ahmed2019, Sahoo2018}, Finslerian gravity \cite{Stavrinos2013}, quintom fields \cite{Feng2006}. Cyclic Universes were considered by Brown et al  in   which phantom scalar fields were used to solve the problem of blackhole overproduction \cite{Brown_2008}.  Another reason for consideration of the cyclic model of the Universe is because they lead to non-inflationary paradigm as they offer an alternative standard inflationary scenario. There are 
%\textbf{Outcome of the work:} Proper dergrees of freedom count, Noether symmetry, dynamical analysis

 In this paper, we  consider the Cuscuton model with a logarithmic extension called the Cucuta-Galileon, coupled to Einstein gravity as in \cite{Rham2017 , Panpanich2023}.  The model is studied in the Friedmann-Lamitre-Robertson-Walker background and  to explore it's  symmetry aspects owing to higher derivative nature. Although one can discard the surface term, but we  consider it's original form so that no information is lost.  With this, the next step is to perform the  Hamiltonian analysis and explore the full constraint structure of the model. The constraint analysis of this model will reveal the true degrees of freedom.  We  show that even with the higher derivative nature of the model it still have only \textit{two} independent degrees of freedom.  We will also investigate the contact symmetries of this model so to determine the conditions for which the model may be invariant under the infinitesimal symmetry transformations.   Additionally, we  will  examine  the Killing symmetries as it is directly connected to the  preservation of the metric structure. 
 
 Another important aspect of this paper is the  investigation the cosmological evolution  of the model. While the authors in \cite{Panpanich2023} have previously explored the dynamical analysis,  our approach focuses on the present model which is subjected to Noether symmetries or more specifically the Lie-Backlund symmetries. Therefore, it will be interesting to see   the dynamics of the various energy densities in the light of the dynamical analysis of the standard cosmology. We will construct the corresponding Friedmann equations and numerically solve the Friedmann equations to study the evolution of the various energy densities and the equation of state parameter over time.  This will shed light on the cosmological implications of the Cuscuton-Galileon extension under symmetry constraints.

The structure of the paper as follows. In Section [\ref{model}] we perform the Hamiltonian analysis of the model from higher derivative point of view and find out the degrees of freedom. In Section [\ref{noether}] we  present the Noether symmetry analysis of the model and find out the infinitesimal parameters. Section [\ref{Killing})] carries out the Killing analysis of the model and find out all the associated charges.  Section [\ref{dynamical}] is devoted see the cosmological dynamical analysis of the model in the minisuperspace framework. Finally, we  conclude our findings in Section [\ref{conclusion}].

\section{The Cuscuton-Gallileon Model of  Gravity}\label{model}
The general action for  the Cuscuton model  was introduced in \cite{Afshordi2007}. Later, this action was modified to accomodate the Galileon fields which is known as the Cuscuta-Galileon model \cite{Maeda2022}. We consider the  action as in \cite{Panpanich2023}, the Cuscuton along with a logarithmic Galileon-term coupled via gravity, is given by
\begin{equation}
	\mathcal{S} = \int \sqrt{-g}  \Big( \frac{M^2}{2} \mathcal{R} + a_2 \sqrt{-X} + a_3 \text{ln}(-\frac{X}{\Lambda^4})\Box \phi + V(\phi)\Big) d^4x.
	\label{action}
\end{equation}
Here $\mc{R}$ represents the Ricci scalar and the $\Box$ is the D'Alembertian operator,  $a_2$ and $a_3$ are some coupling constants.  We also consider $sgn(\dot{\phi})$ as positive to reduce any ambiguation in calculations.
Let us consider the background metric as FRLW metric, wtih $a(t)$ as scale factor,  given by 
\begin{equation}
	ds^2 = -dt^2 + a(t)^2 \delta_{ij}dx^i dx^j.
	\label{frlw_metric}
\end{equation}
Applying the metric (\ref{frlw_metric})  in terms of the fields, the action (\ref{action})  becomes 
\begin{eqnarray}
	\mathcal{S}  
	= \int d^4x \Big( 3M^2 (a \dot{a}^2 + a^2\ddot{a}) + a_2 a^3 \dot{\phi} - a_3 \ln \frac{\dot{\phi}^2}{\Lambda^4} (a^3 \ddot{\phi} + 3 \dot{a} \dot{\phi}) - a^3 V(\phi)\Big).
	\label{lag1}
\end{eqnarray} 
An overdot $` \ \dot{}$ \  ' represents derivative with respect to the timelike coordinate $t$. The Lagrangian in this case is a  higher derivative model. Redefining the fields as $A=\dot{a}, \Phi = \dot{\phi}$ and introducing two new fields as Lagrange multipliers $\lambda_a, \lambda_\phi$ we get the first order Lagrangian as 
\begin{eqnarray}
	\mathcal{L'} = 3M^2 (a A^2 + a^2\dot{A}) + a_2 a^3 \Phi - a_3 \ln \frac{\Phi^2}{\Lambda^4} (a^3 \dot{\Phi} + 3 A \Phi) - a^3 V(\phi) + \lambda_a(A - \dot{a}) + \lambda_\phi(\Phi - \dot{\phi}).
	\label{lag2}
\end{eqnarray}
By this redefinition of the fields the new Lagrangian (\ref{lag2}) is now first order in form. Introduction of the Lagrange multipliers is necessary to accommodate new degrees of freedom and this indeed indeed expand the dimension of the configuration space.  We now perform the Hamiltonian analysis of the model.
\subsection{Hamiltonian analysis }

The phase space can be constructed from this Lagrangian (\ref{lag2}) which is spanned by  the fields $q = \{ a, A, \phi, \Phi , \lambda_a, \lambda_\phi \}$ and their corresponding momenta $P = \{ P_a, P_A, P_\phi, P_\Phi, P_{\lambda_a}, P_{\lambda_\phi}\}$. The conjugate momenta by usual definition $P = \frac{\p \mc{L}'}{\p \dot{q}},$ are found as
\begin{eqnarray}
	P_a = - \lambda_a, \ \  P_A = 3M^2a^2, \ \   P_\phi =-\lambda_\phi, \ \  
	P_\Phi = -a_3 a^3 \ln \frac{\Phi^2}{\Lambda^4}, P_{\lambda_a}=0, P_{\lambda_\phi} =0  
	\label{momenta}
\end{eqnarray}
As the momenta are  not invertible in terms of the velocities, we find the primary constraints as, 
\begin{eqnarray}
	\Psi_1 = P_a + \lambda_a \approx 0, \ \ \Psi_2 = P_\phi + \lambda_\phi \approx 0, \ \  \Psi_3 = P_{\lambda_a} \approx 0, \Psi_4 = P_{\lambda_\phi} \approx 0 , \\ 
	\Psi_5 = P_A - 3M^2 a^2 \approx 0, \ \ \  \Psi_6 = P_\Phi + a_3 a^3 \ln \frac{\Phi^2}{\Lambda^4} \approx 0.
	\label{primary}
\end{eqnarray}
The non-zero Poission brackets between the primary constraints are given below 
\begin{eqnarray}
	\{ \Psi_1, \Psi_3\} =1, \ \ \{ \Psi_1, \Psi_5\} = 6M^2 a, \ \  \{\Psi_1, \Psi_6 \} = -3 a_3 a^2 \ln \frac{\Phi^2}{\lambda^4}, \ \  \{ \Psi_2, \Psi_4\} = 1.
\end{eqnarray}
We obtain the canonical Hamiltonian as 
\begin{eqnarray}
	H_\text{can} = -3 M^2 a A^2 - a_2 a^3 \Phi + 3a_3 A \Phi \ln \frac{\Phi^2}{\Lambda^4} + P_a A + P_\phi \Phi + a^3 V(\phi)+\sum_{i=1,6} \Lambda_i \Psi_i  
\end{eqnarray}
Once we have found out the primary constraints, it is customary to check  the time evolution of these constraints by considering Poission brackets with the canonical Hamiltonian 	as $\dot{\Psi_i} = \{ \Psi_i, H_{can}\}$. Thus, we obtain 
\begin{eqnarray}
	\dot{\Psi_1} &=&  -3 a_3 a^2 \Lambda _6 \ln \left(\frac{\Phi ^2}{\Lambda ^4}\right)-3 a^2 V(\phi )+3 a_2 a^2 \Phi +6 a \Lambda _5 M^2+3 A^2 M^2+\Lambda _3, \label{const_eom1}\\ 
	\dot{\Psi_2}  &=&  \Lambda _4-a^3 V'(\phi ), \label{const_eom2}\\ 
	\dot{\Psi_3}  &=& \Lambda_1, \label{const_eom3}\\ 
	\dot{\Psi_4}  &=&  \Lambda_2, \label{const_eom4} \\ 
	\dot{\Psi_5}  &=& -6 a M^2 \left(A+\Lambda _1\right)+6 a A M^2-3 a_3 \Phi  \ln \left(\frac{\Phi ^2}{\Lambda ^4}\right)-P_a, \label{const_eom5}\\ 
	\dot{\Psi_6} &=&   a_2 a^3+3 a_3 a^2 \left(A+\Lambda _1\right) \ln \left(\frac{\Phi ^2}{\Lambda ^4}\right)-3 a_3 A \ln \left(\frac{\Phi ^2}{\Lambda ^4}\right)-6 a_3 A-P_{\phi }. \label{const_eom6} \\ 
\end{eqnarray}
Equating these time preservation to zero, we get $\Lambda_1 = \Lambda_2 =0, \Lambda_4 = a^3 V'(\phi)$. Here a prime $'$ in $V'(\phi)$ refers to the first derivative with respect to $\phi$. Hence   two new secondary constraints from (\ref{const_eom5}) and (\ref{const_eom6}) emerge as
\begin{eqnarray}
 && \chi_1 =  3 a_3 \Phi  \ln \left(\frac{\Phi ^2}{\Lambda ^4}\right)  + P_a \approx 0, \label{second_const_1}\\ 
 && \chi_2 =  a_2 a^3+3 a_3( a^2-1) A \ln \left(\frac{\Phi ^2}{\Lambda ^4}\right) -6 a_3 A-P_{\phi } \approx 0 .\label{second_const_2}
\end{eqnarray}
The Poission Brakets between the constraints are given by 
\begin{eqnarray}
&&	\{\chi_1, \Psi_5 \} = 6 a M^2 , 
	\{\xi_1, \Psi_6 \} = 3 a_3 \left(2-\left(a^2-1\right) \ln \left(\frac{\Phi ^2}{\Lambda ^4}\right)\right)\\
	&& \{\chi_2, \Psi_1 \}  = 3 a \left(2 a_3 A \ln \left(\frac{\Phi ^2}{\Lambda ^4}\right)+a a_2\right)  ,\\
	&& \{\chi_2, \Psi_5 \}  =   3 a_3 \left(\left(a^2-1\right) \ln \left(\frac{\Phi ^2}{\Lambda ^4}\right)-2\right) ,
	 \{\xi_2, \Psi_6 \} =  \frac{6 \left(a^2-1\right) a_3 A}{\Phi } \\
&& \{\chi_1, \chi_2 \} = -3 a_2 a^2-6 a_3 a A \ln \left(\frac{\Phi ^2}{\Lambda ^4}\right).
\end{eqnarray}
Time preservation of the secondary constraints lead to 
\begin{eqnarray}
	\dot{\chi_1} &=& 3 \left(a_3 \Lambda _6 \left(2-\left(a^2-1\right) \ln \left(\frac{\Phi ^2}{\Lambda ^4}\right)\right)-a^2 V(\phi )+a_2 a^2 \Phi +2 a \Lambda _5 M^2+A^2 M^2\right) , \label{const_eom7} \\
	\dot{\chi_2}& =&   a^3 V'(\phi )+\frac{6 \left(a^2-1\right) a_3 A \Lambda _6}{\Phi }+3 a_3 \Lambda _5 \left(\left(a^2-1\right) \ln \left(\frac{\Phi ^2}{\Lambda ^4}\right)-2\right) \nonumber \\ && +3 a \left(A+\Lambda _1\right) \left(2 a_3 A \ln \left(\frac{\Phi ^2}{\Lambda ^4}\right)+a a_2\right) \label{const_eom8}.
\end{eqnarray}

Equating the equations  (\ref{const_eom7}, \ref{const_eom8})  and  (\ref{const_eom1}) to zero and solving  we get 
\begin{eqnarray}
	\nonumber
	 \Lambda_5& =&\frac{1}{\Xi_1}\Big(  (2-\left(a^2-1\right) \ln \left(\frac{\Phi ^2}{\Lambda ^4}\right)) (a^3 V'(\phi )+3 a A (2 a_3 A \ln \left(\frac{\Phi ^2}{\Lambda ^4}\right)+a a_2)) \nonumber \\ && -\frac{6}{\Phi} \left(a^2-1\right) A \left(a_2 a^2 \Phi +A^2 M^2-a^2 V(\phi ) \right) \Big) , 
	 \nonumber \\   
	  \Lambda_6 &=&\frac{\Phi}{\Xi_2} \Big( 6 a^2 a_3 V(\phi ) -2 a^4 M^2 V'(\phi )-6 a_2 a^3 A M^2-6 a_2 a_3 a^2 \Phi -6 a_3 A^2 M^2 \nonumber \\  &&
	  	-3 \ln \left(\frac{\Phi ^2}{\Lambda ^4}\right)( a_3 A^2 M^2 +3 a^2 a_3 A^2 M^2 + a^2 a_2 a_3 \Phi  -a^4 a_2 a_3 \Phi  - a^2a_3\left(a^2 -1 \right) )\Big).
\end{eqnarray}
Here we have considered 
\begin{eqnarray}
	\nonumber 
	\Xi_1 &=& 3 \left(\frac{4 a \left(a^2-1\right) A M^2}{\Phi }+a_3 \left(\left(a^2-1\right) \ln \left(\frac{\Phi ^2}{\Lambda ^4}\right)-2\right)^2\right),
	\\
	\nonumber
	\Xi_2&=& 3 a_3 \left(4 \left(a^3 A M^2-a A M^2+a_3 \Phi \right)+\left(a^2-1\right)^2 a_3 \Phi  \ln ^2\left(\frac{\Phi ^2}{\Lambda ^4}\right)-4 \left(a^2-1\right) a_3 \Phi  \ln \left(\frac{\Phi ^2}{\Lambda ^4}\right)\right).
\end{eqnarray}
In this way $\Lambda_3$ also can be found out by solving (\ref{const_eom1}) easily.

\subsection{DOF count}                                                                                           
In the HD form of this theory given by (\ref{lag2}),there are a total of 8 second class constraint. The phasespace is 12 dimensional. Hence the no of degrees of freedom is 
 $ \frac{1}{2}(\text {dim. of phase space } - 2 \times \text {no of first class constraints}  - \text {no of second class constraints})  = 2, $
which  agrees with the results found in \cite{Afshordi2007, Gomes2017, Mukohyama2019}.   The model in (\ref{lag2}) can be written explicitly in a form so that it has a total derivative term. One can integrate out the surface term and  identify the two degress of freedom which are $a(t)$ and $\phi(t)$.  In the next section, we will get rid of this surface term and determine the Noether symmetry parameters.

\section{The integrated out Lagrangian }\label{noether}
% The Lagrangian can be written as 
 %\begin{eqnarray}
 	%\mathcal{L} = -3M^2 a \dot{a}^2 + a_2 a^3\dot{\phi} + 6 %a^2a_3\dot{a}\dot{\phi} - a^3 V(\phi)
% \end{eqnarray}
Now, to explore the Noether symmetries of the system, we  write the HD lagrangian  after identifying the surface terms in the following, as
\begin{eqnarray}
	\mathcal{L} = -3M^2 a \dot{a}^2 - 3a_2 a^2\dot{a}\phi + 6 a^2a_3\dot{a}\dot{\phi} - a^3 V(\phi) + \frac{d}{dt}(a_2a^3\phi).
	\label{surface_lag}
\end{eqnarray}
It is interesting to see a term which is total derivative in nature and we have kept this term as it will be helpful to fix  the Noether symmetries.
\subsection{Noether symmetries}
Consider a one-parameter   infinitesimal point  transformation of  the fields in the form 
\begin{eqnarray}
&&	t \rightarrow t' = t + \epsilon \eta_t(t,a(t), \phi(t)) \\
	&& a(t) \rightarrow a'(t) = a + \epsilon \eta_a(t,a(t), \phi(t)) \\
	&& \phi(t) \rightarrow \phi'(t) = \phi + \epsilon \eta_\phi(t,a(t), \phi(t)),
\end{eqnarray}
we can calculate the variables $\eta_t,\eta_a, \eta_\phi$ that keep the contact symmetries intact.  Here $\epsilon$ is the infinitesimal parameter.
We can construct the generator from these infinitesimal parameters as
\begin{equation}
X = \eta_t\frac{\partial}{\partial t} + \eta_a \frac{\partial}{\partial a} + \eta_\phi \frac{\partial}{\partial \phi}.
\end{equation}
 
Then the first prolongation of the generator $X(\eta_t, \eta_a, \eta_\phi)$ can be written as \cite{ BlumanB}
\begin{equation}
	X^{[1]}(\eta_t, \eta_a, \eta_\phi) = \eta_t \partial_t + \eta_a \partial_a + \eta_\phi \partial_\phi + (\dot{\eta}_a - \dot{a}\dot{\eta}_t)\partial_{\dot{a}}  + + (\dot{\eta}_\phi - \dot{\phi}\dot{\eta}_t)\partial_{\dot{\phi}} 
	\end{equation}
 The first prolongation of the infinitesimal generator obeys \cite{BlumanB, Paliathanasis2016a}
\begin{eqnarray}
	X^{[1]}\mathcal{L}_0 + \mathcal{L}_0 \dot{\eta}_t = \dot{\sigma}.
	\label{prolong1}
\end{eqnarray}
Where we have identified a surface term as $\sigma = a_2a^3\phi$ and $\mathcal{L}_0$ is the part from the Lagrangian without the surface term in (\ref{surface_lag}) .  
 Now we apply (\ref{prolong1}) and equate the various coefficients of the derivatives of $a(t)$ and $\phi(t)$. After a straightforward comparison  we get the following sets of differential equations 
\begin{eqnarray}
 	-3a_2\eta_\phi a^2 - 6M^2a\partial_t\eta_a + 6 a_3 a^2\partial_t\eta_\phi -a^3\partial_a\eta_t V(\phi) - 6a_2\eta_aa\phi -3a_2 a^2 \phi  - 3a_2 a^2 \phi \partial_a \eta_a =0 \label{coeef_eqn1}\\
 -3M^2 \eta_a - 6 M^2a\partial_a\eta_a + 6 a_3a^2\partial_a\eta_\phi + 3M^2a\partial_t\eta_t =0 \label{coeef_eqn2}\\
 3M^2 a \partial_a \eta_t =0 \label{coeef_eqn3}\\
 	-a_2 a^3 + 6 a_3 a^2 \partial_t \eta_a  - a^3\partial_\phi\eta_t V(\phi) -3a_2a^2\partial_\phi\eta_a\phi =0 \label{coeef_eqn4}\\
 	12a_3\eta_a a+6a_3a^2\partial_a\eta_a -6M^2a\partial_\phi \eta_a + 6a_3a^2\partial_\phi\eta_\phi -6a_3a^2\partial_t\eta_t =0 \label{coeef_eqn5}\\
 	-6a_3a^2\partial_a \eta_t + 3M^2a\partial_\phi \eta_t =0 \label{coeef_eqn6}\\
 	6a_3 a^2 \partial_\phi\eta_a =0 \label{coeef_eqn7}\\
 	6a_3a^2\partial_\phi \eta_t= 0 \label{coeef_eqn8}\\
 	-3a^2V(\phi)\eta_a-a^3\partial_t\eta_t V(\phi) -3a_2a^2\partial_t\eta_a \phi - \eta_\phi a^3\partial_\phi V(\phi) =0. \label{coeef_eqn9}
\end{eqnarray}
We now solve the partial differential equations (\ref{coeef_eqn1} - \ref{coeef_eqn9}).  Using (\ref{coeef_eqn3}) and (\ref{coeef_eqn8}) we infer that $\eta_t$ is function of $t$ only. Therefore, we get 
\begin{equation}
	\eta_t = ct+c_1.
\end{equation}
Where $c$ and $c_1$ are arbitrary constants. Now, it is imperative that we can assume $\eta_a$ and $\eta_\phi$ are functions of only $a$ and $\phi$ respectively, we solve (\ref{coeef_eqn2}, \ref{coeef_eqn5}) using separation of variable method and obtain  
\begin{eqnarray}
	  \eta_a = ac/3,  \eta_\phi = c, V(\phi ) = k e^{-2\phi}. 
\end{eqnarray}
Next  we consider (\ref{coeef_eqn4})  which yield $a_2=0$ and thus (\ref{coeef_eqn1}) become trivial. Imposing these results on (\ref{coeef_eqn9})  we get $V(\phi) = k e^{-2 \phi}$.  It is trivial to check that for these values of the coefficients $a_2, \eta_t, \eta_a, \eta_\phi$ the equation (\ref{prolong1}) is satisfied identically.

  This result is interesting as it comes out that in order to preserve the contact symmetry of the system we need an exponential potential and the above system should have $a_2 =0$. This changes scenario as the non-zero value of the coefficient $a_2$ in Cuscuton gravity breaks the contact symmetry of the system. This is also evident from the infinite sound speed and the non-local characteristics of the Cuscuton gravity. However, now we shall consider the model with $a_2 =0$ i.e. without the original Cuscuton term and explore  the effect of the other logarithmic term from the point of view of the cosmological evolution of the model. 

\section{Killing analysis in the minisuperspace version}\label{Killing}
Now we consider the Lagrangian (\ref{surface_lag})  obeying the contact symmetries and for that we put $a_2 =0$. Therefore, the Lagrangian become 
\begin{eqnarray}
	\nonumber
	\mathcal{L} &&=  -3M^2 a \dot{a}^2 + 6 a_3 a^2 \dot{a}\dot{\phi} - a^3V(\phi)  \label{lag3}\\
	&&=  \frac{1}{2}\gamma_{ij}\zeta^i \zeta^j - a^3 V(\phi). \label{mini_lag}
\end{eqnarray} 
Now, in the mini-superspace version of the theory we can identify the  space spanned by the fields $\zeta = \{a(t),\phi(t)\} $. Here $\{i,j\}$ are the indices corresponding to the metric $\gamma_{ij}$ in  (\ref{mini_lag}).  The metric over which the space is constructed is called kinetic metric given by
\begin{eqnarray}
	\gamma_{ij} = \begin{pmatrix}
		-6M^2 a & 6a_3a^2 \\ 6a_3a^2 & 0
	\end{pmatrix}.
\end{eqnarray}
  
This is actually a flat metric with the non-zero Christoffels given by
\begin{equation}
	\Gamma^{a}_{aa} = \frac{2}{a}, \Gamma^{\phi}_{aa} =  \frac{3M^2}{a_3a^2}.
\end{equation}
 
The phasespace is now spanned by $\{ a(t), \phi(t), \Pi_a(t), \Pi_\phi(t)\}$ for this mini-superspace Lagrangian (\ref{mini_lag}). We  calculate the momenta $\{ \Pi_a(t), \Pi_\phi(t)\}$ corresponding to the fields $a(t)$ and $\phi(t)$ as
\begin{eqnarray}
	\Pi_{a} &=& \frac{\partial \mathcal{L}}{\p \dot{a}} =  6 a_3 a^2 \dot{\phi}-6 M^2 a \dot{a}, \\
	\Pi_{\phi} &=&\frac{\partial \mathcal{L}}{\p \dot{\phi}} =  6 a_3 a^2 \dot{a}.
\end{eqnarray}
All the velocities are invertible in terms of the momenta and therefore there appears no constraints. Thus the constraint structure of the system has also been been changed due to the consideration of the contact symmetries of the model.  The Hamiltonian can be found as
\begin{eqnarray}
	H = \frac{M^2 \Pi _{\phi }{}^2}{12 a_3^2 a^3}+\frac{\Pi _a \Pi _{\phi }}{6 a_3 a^2}+a^3 V(\phi ).
	\label{mini_ham}
\end{eqnarray} 
The Hamiltonian will be useful for time evolution of the dynamical quantities.

\subsection{Killing vectors and tensors of the model}
In this subsection we shall consider the diffeomorphism symmetry of the metric which essentially is nothing but the Killing symmetries. As a consequence, the Killing vectors $\xi_i$ corresponding to the Kinetic matrix $\gamma_{ij}$ defined above obey the Killing equations \cite{Carrol2019}
\begin{eqnarray}
	\nabla_i\xi_j + \nabla_j \xi_i =0. \label{killingeqns}
\end{eqnarray}
The three Killing vectors are obtained by solving the Killing equations (\ref{killingeqns}) as
\begin{eqnarray}
\xi_{(1)} &&= 
  \frac{1}{6a_3}\partial_{\phi} ,\\
\xi_{(2)} &&= 
	\frac{1}{6 a^2 a_3}\partial_{a} + \frac{M^2}{12 a^3 a_3^2}\partial_{\phi}, \\  
\xi_{(3)}&& = 
\frac{a}{18  a_3} \partial_{a} + \frac{1}{36  a_3^2}\Big( M^2 (3 \ln a+2)-6 a_3 \phi \Big)\partial_{\phi}.
\end{eqnarray}
We can obtain the Killing tensors out of these defined as 
\begin{equation}
	 K^{ij} = \frac{1}{2}K^{(i}\otimes K^{j)}.
\end{equation} 
Therefore, the Killing tensors we obtain as 
\begin{eqnarray}
	K^{ij}_{(11)}=\left(
	\begin{array}{cc}
		0 & 0 \\
		0 & \frac{1}{36  a_3^2} \\
	\end{array}
	\right), 
	K^{ij}_{(12)} =\left(
	\begin{array}{cc}
		0 & \frac{1}{72 a^2 a_3^2} \\
		\frac{1}{72 a^2 a_3^2} & \frac{M^2}{72 a^3 a_3^3} \\
	\end{array}
	\right)\\
	K^{ij}_{(13)} = \left(
	\begin{array}{cc}
		0 & \frac{1}{5216 a_3^2} \\
		\frac{1}{216 a_3^2} & -\frac{\Sigma_1}{216 a_3^3} \\
	\end{array}
	\right), 
	K^{ij}_{(22)} =\left(
	\begin{array}{cc}
		\frac{1}{36a^4 a_3^2} & \frac{M^2}{72 a^{5} a_3^3} \\
		\frac{M^2}{9 a^{5} a_3^3} & \frac{M^4}{144 a^{6} a_3^4} \\
	\end{array}
	\right), \\ 
	K^{ij}_{(23)} = \left(
	\begin{array}{cc}
		\frac{1}{108 a a_3^2} & -\frac{\Sigma_2}{144 a^2 a_3^3} \\
		-\frac{\Sigma_2}{144 a^2 a_3^3} & -\frac{M^2\Sigma_1 }{432 a^3 a_3^4} \\ 
	\end{array}
	\right), 
	K^{ij}_{(33)} = \left(
	\begin{array}{cc}
		\frac{a^2}{324  a_3^2} & \frac{a \Sigma_1}{ 648 a_3^3 } \\
		\frac{a \Sigma_1}{648  a_3^3} & \frac{\Sigma_1^2}{1296  a_3^4} \\
	\end{array}
	\right)
\end{eqnarray}
where we have considered $\Sigma_1 = M^2 (3 \ln a+2)-6 a_3 \phi, \Sigma_2 = M^2 (\ln a+1)-2 a_3 \phi  $. 

We can calculate the charges corresponding to the Killing vectors $\xi^1, \xi^2$ and $\xi^3$ using \cite{Carrol2019}
\begin{eqnarray}
	Q_{(a)} =  \gamma_{ij}\xi_{(a)}^{i} \frac{d\zeta^j}{dt}.
\end{eqnarray}
Which, in terms of the phasespace variables, are found as
\begin{eqnarray}
	Q_{(1)} &=&  \frac{\Pi _{\phi }}{6 a_3}, \\ 
	Q_{(2)} &=&  \frac{1}{12 a_3^2 a^3}  \Big( 2 a_3 a \Pi _a+M^2 \Pi _{\phi }\Big),\\
	Q_{(3)} &=&  \frac{1}{36 a_3^2}\Big(  M^2 \Pi _{\phi } (3 \ln (a)+2)-6 a_3 \phi \Pi _{\phi } +2 a_3 a \Pi _a \Big).
\end{eqnarray}

Also the charges corresponding to the Killing-Tensors can be found out as
\begin{eqnarray}
	Q_{(ab)} =  \gamma_{ik}\gamma_{jl}  K_{(ab)}^{ij}\frac{d\zeta^k}{dt} \frac{d\zeta^l}{dt}.
\end{eqnarray}
The charges corresponding to the Killing-Vectors are found out to be
\begin{eqnarray}
	Q_{(11)} &=& \frac{\Pi _{\phi }{}^2}{36 a_3^2}, \\ 
	Q_{(12)} &=&  \frac{1}{72 a_3^3 a^3}   \left(2 a_3 a \Pi _a\Pi _{\phi }+M^2 \Pi _{\phi }^2\right), \\
	Q_{(13)} &=& \frac{1}{216 a_3^3}\Big(  M^2 \Pi _{\phi }^2 (3 \ln (a)+2)-6 a_3 \phi \Pi _{\phi }^2 + 2 a_3 a \Pi _a \Pi _{\phi }\Big), \\
	Q_{(22)} &=& \frac{1}{144 a_3^4 a^6} \left(2 a_3 a \Pi _a +M^2 \Pi _{\phi } \right){}^2,\\
	Q_{(23)} &=& \frac{1}{432 a_3^4 a^3} \left(2 a_3 a \Pi _a +M^2 \Pi _{\phi }\right) \left( M^2 \Pi _{\phi }(3 \ln (a)+2)-6 a_3 \phi \Pi _{\phi } +2 a_3 a \Pi _a \right), \\
	Q_{(33)} &=& \frac{1}{1296 a_3^4}\left( M^2 \Pi _{\phi }(3 \ln (a)+2)-6 a_3 \phi \Pi _{\phi }+2 a_3 a \Pi _a \right){}^2. \\
\end{eqnarray}
The charges corresponding to the Killing vectors and Killing tensors thus calculated are actually the conserved quantities. To ensure conservation of the charges, thus obtained, we calculate the preservation over time  as 
\begin{eqnarray}
	\{Q_{(a)}, H\} = \{ Q_{(ab)} , H\} =0.
\end{eqnarray} 
This indicate that the charges are trully constants of motion. We will next see the cosmological evolution of the model after considering $a_2 =0$.
\section{Dynamical system analysis}\label{dynamical}
The equations of motion for the Lagrangian \ref{lag3} is given by
\begin{eqnarray}
	3M^2 H^2 - \rho_m -\rho_r - V(\phi) - 12 a_3 H\dot{\phi} = 0, \\ 
	3M^2H^2 + 2M^2\dot{H} + P_m + P_r - V(\phi) + 2 a_3 \ddot{\phi} =0, \\
	18a_3 H^2 + 6a_3 \dot{H} - V'(\phi) =0.
\end{eqnarray} 
We define the dimensionless dynamical variables as
\begin{eqnarray}
	x = \frac{V(\phi)}{3M^2H^2}, y = \frac{2a_3 \dot{\phi}}{M^2H}, z= \frac{\rho_r}{3M^2H^2}.
\end{eqnarray}

Using these dimensionless variables we obtain the set of dynamical equations by differentiating each variable by $N = \text{ln}a$.
Using these of dynamical equations we get the corresponding equations of motions as
\begin{eqnarray}
	\frac{\dot{H}}{H^2} = -3 + \frac{1}{2}\lambda x, \\
	\frac{dx}{dN} = \frac{1}{2}\lambda x y - 2 x \frac{\dot{H}}{H^2}, \label{dynamical1} \\
	\frac{dy}{dN} = 3-\lambda x - z + 3 x - y \frac{\dot{H}}{H^2},
	\label{dynamical2} \\
	\frac{dz}{dN} = -4z - 2z \frac{\dot{H}}{H^2}.
	\label{dynamical3}
\end{eqnarray}
 The effective equation of state is given by
\begin{eqnarray}
	\omega_{eff} = -1-\frac{\dot{2H}}{3H^2} = 1- \frac{1}{3}\lambda x .
	\label{omega_eff}
\end{eqnarray}
Using these definitions we can easily write down the equations of state
\begin{eqnarray}
	\omega_\phi = \frac{\omega_{eff} - \frac{1}{3}z}{x+2y}. 
	\label{omega_phi}
\end{eqnarray}
The absolute energy density parameters corresponding  to radiation, matter and the field $\phi$,  the definitions of the dynamical variables $x,y$ and $z$, are given by  
\begin{eqnarray}
	\Omega_r &=& z, \\
	\Omega_\phi &=& x+ 2y, \\
	\Omega_m &=& 1 - (\Omega_\phi + \Omega_r).
\end{eqnarray} 
In order to find the stability of the dynamical equations we need to find out the fixed points of the system.  By equating the R.H.S. of the dynamical equations (\ref{dynamical1}- \ref{dynamical1}) to zero we get the fixed points as in Table (\ref{table1})

\begin{tabular}{|c|c|c|c|c|c|c|c|}
	\hline 
	Fixed points & x& y& z&$\Omega_r$& $\Omega_m$&$\Omega_\phi$ & Conditions \\ 
	\hline 
	A  &0 & -1& 0&0 &3 &-2&Unstable \\
	\hline 
	B&$\frac{2}{\lambda}$ & $\frac{-8}{\lambda}$ & $\frac{\lambda - 10}{\lambda}$ &  $\frac{\lambda - 10}{\lambda}$& $\frac{24}{\lambda }$ &- $ \frac{14}{\lambda}$&Unstable \\ 
	\hline 
	C&$\frac{3}{\lambda}$& -$\frac{6}{\lambda}$& 0 & 0& $\frac{\lambda +9}{\lambda }$&-$\frac{9}{\lambda}$&stable  $\lambda >9$ \\
	\hline 
	D & -1 + $\frac{12}{\lambda}$& -2+$\frac{12}{\lambda}$& 0 &  0& $\frac{6 (\lambda -6)}{\lambda }$& 5+ $\frac{36}{\lambda}$&stable for $\lambda <9$ \\
	\hline %dynamical_analysis.nb%
	
\end{tabular}\label{table1}

We now examine   the behavior of the dynamical variables around these fixed points A,B, C and D. For that we expand the dynamical equations (\ref{dynamical1}-\ref{dynamical3}) to first order for the variables $\vec{X} = {x,y,z}$ we get the variation of the transformations in the form of $\frac{d}{dN}\delta \vec{X} = \mathcal{M} \delta \vec{X}$ where we have obtained the transformation matrix as 
\begin{eqnarray}
	\mathcal{M} = \begin{pmatrix}
		6-2  \lambda x +\frac{\lambda  y}{2} & \frac{\lambda  x}{2} & 0\\
		3-\frac{1}{2} \lambda  (y+2) & 3-\frac{\lambda  x}{2} & -1 \\
		-z \lambda &  0 & 2-\lambda  x
	\end{pmatrix}.
	\label{jacobian_Matrix}
\end{eqnarray}
Below we point out the behavior as can be inferred form Matrix $\mc{M}$ for the fixed points A, B, C and D.

\begin{figure}
	\includegraphics[scale=0.5]{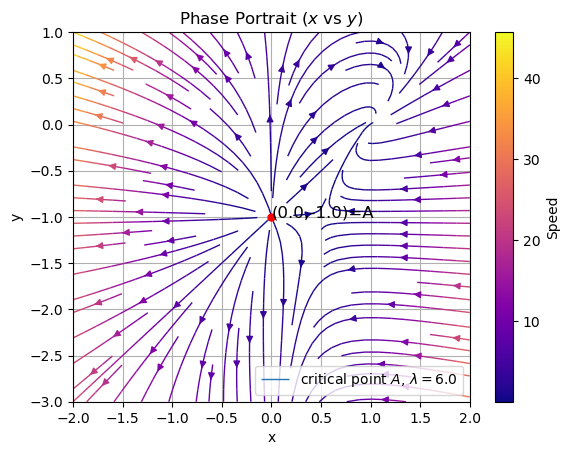}
		\includegraphics[scale=0.5]{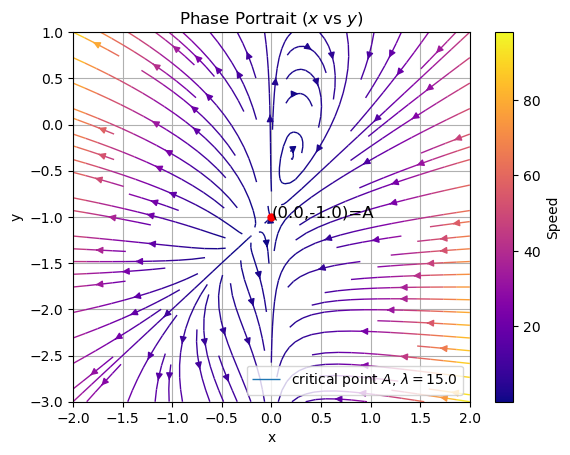}
 	\caption{Figure for evolution of of x vs y,   for the critical point A with $\lambda <12$ and $\lambda>12$.}
	\label{critical_A}
\end{figure}

\textbf{Critical point A:} For the critical point A($0,-1,0$) one can see that it is also not dependent on $\lambda$ and hence we can say that it is not dependent on the potential. The eigen values of the Jacobian matrix (\ref{jacobian_Matrix}) are  $\left\{3,2,6-\frac{\lambda }{2}\right\}$.  It is unstable a point for $\lambda < 12$ as all the eigenvalues are positive. For $\lambda > 12$ it is a  saddle point with 1 dimensional stable and 2 dimensional unstable manifold. For this point, at the beginning the radiation density is zero with the -ve scalar field energy density. Due to this, matter energy density $\Omega_r$ turns out to be greater that 1 which is unfeasible for the evolution of the universe. The unstable behavior of the phase-space variables are shown by projecting them on a 2D plane for the critical point A in Figure (\ref{critical_A}).

\begin{figure}
	\includegraphics[scale=0.35]{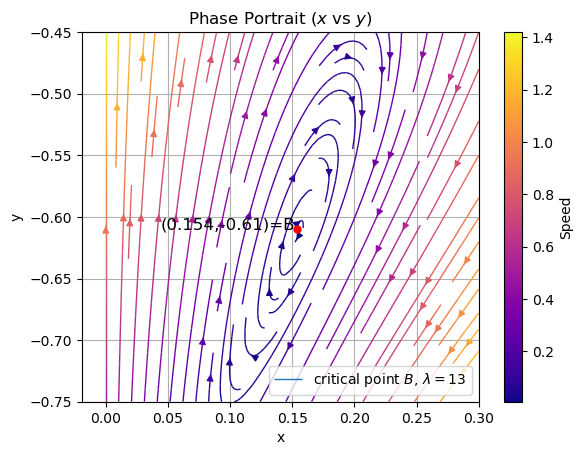}
	\includegraphics[scale=0.35]{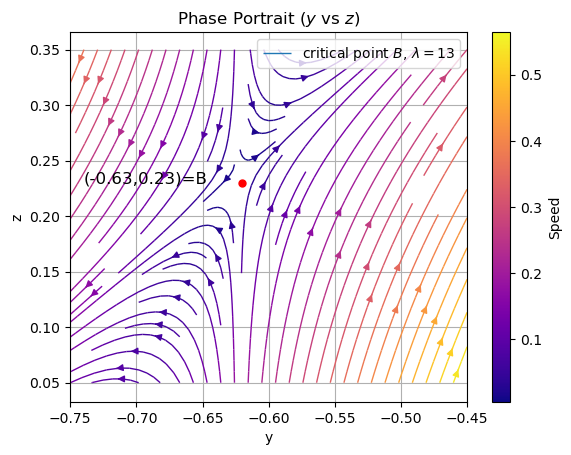}
	\includegraphics[scale=0.35]{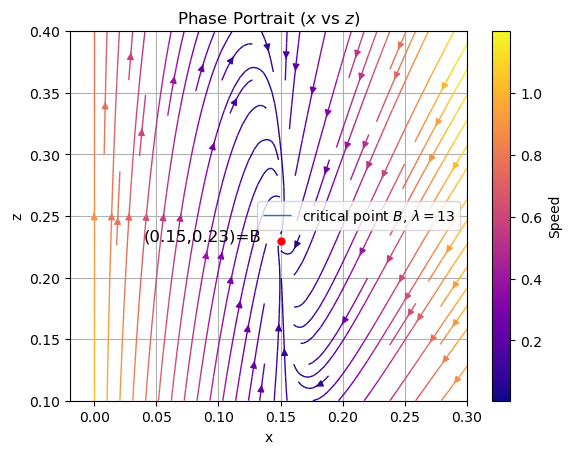}
	\caption{Figure for evolution of  $x-y$, $y-z$ and $x-z$ for the critical point B with $\lambda =13$.}
	\label{image_B}
\end{figure}

\textbf{Critical point B:}
In this case, the critical point is ($\frac{2}{\lambda}, - \frac{8}{\lambda}, 1-\frac{10}{\lambda}$) and the Jacobian matrix has  eigenvalues $\left\{1,\frac{1}{2} \left(-\sqrt{41-4 \lambda }-1\right), \frac{1}{2}  \left(\sqrt{41-4 \lambda }-1\right)\right\}$. Clearly the eigenvalues are dependent on the values of $\lambda$. Three cases arise: (a) for $\lambda > 41/4$ it has 1 real +ve and 2 imaginary eigenvalues with -ve real part, (b) for $10<\lambda<41/4$ there appear 1 +ve and 2 -ve real eigen values (c) for $\lambda < 10$ there are 2 +ve and 1 -ve real eigen values. The scenario for case (a) is indicated in Fig. (\ref{image_B}). At the beginning the radiation density is zero only for $\lambda =0$. It is +ve for $\lambda<0$ or $\lambda>10$ and -ve for $0<\lambda<10$.  As indicated from the table (\ref{table1}) it is clear that the desirable values are $\lambda>24$ for the critical point B   so that $\Omega_m >1$.   
 \begin{figure}
 	\includegraphics[scale=0.5]{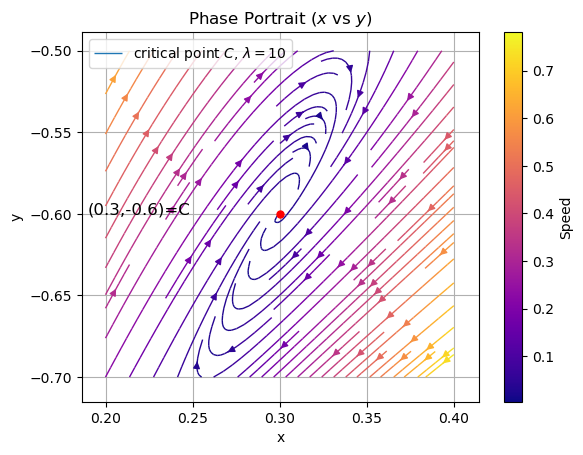}
 	\includegraphics[scale=0.5]{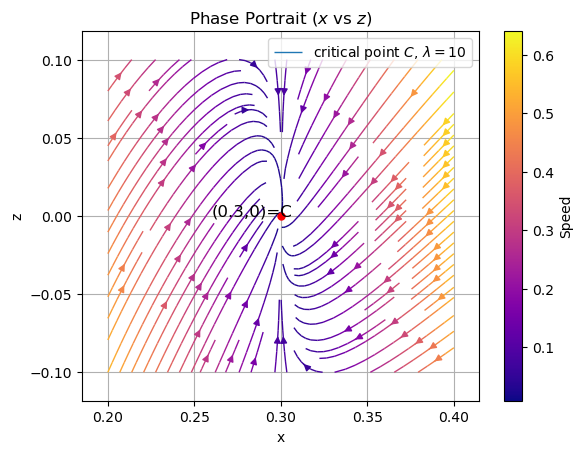}
 	\caption{ Case (e) for the critical point C with $\lambda  > 75/8$. Evolution of  x vs y, y vs z and x vs z for the critical point B with $\lambda =13$.}
 	\label{image_C}
 
 \end{figure}
 
\textbf{Critical point C:}
This interesting critical point is situated at ($3/\lambda, -6/\lambda, 0$).  The jacobian matrix has eigen values $ \left\{-1,\frac{1}{4} \left(- \sqrt{3(75-8 \lambda )}-3\right),\frac{1}{4} \left( \sqrt{3(75- 8 \lambda) }-3\right)\right\}$. Depending on the values of $\lambda$ there can be three situations (c) a stable node for $\lambda = 75/8$ as all three eigenvalues are negative (d) for $\lambda < 75/8$  there appear two -ve and one +ve real eigen values which give a stable node on the x-y subspace and (e) for $\lambda > 75/8$ there appear one -ve and two complex eigenvalues with -ve real parts have a stable spiral on the y-z subspace. At the early times the radiation density is zero for all values of $\lambda$. However, A positive $\lambda$ always give rise to negative energy density of the scalar fields and a positive matter density which is greater than 1 and undesirable.   

\begin{figure}
	\includegraphics[scale=0.35]{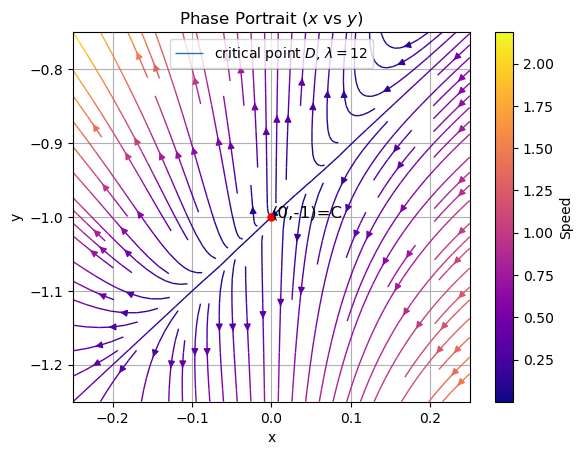}
	\includegraphics[scale=0.35]{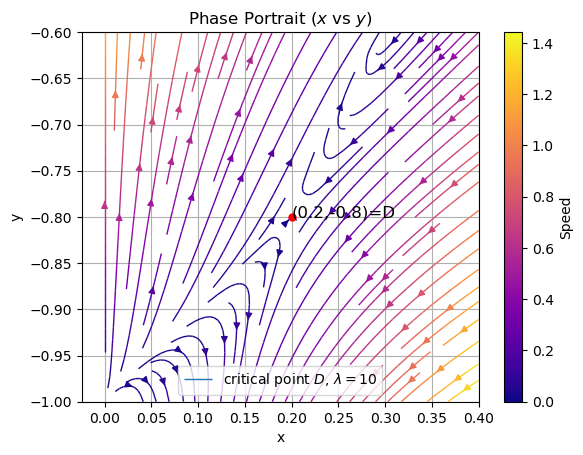}
	\includegraphics[scale=0.35]{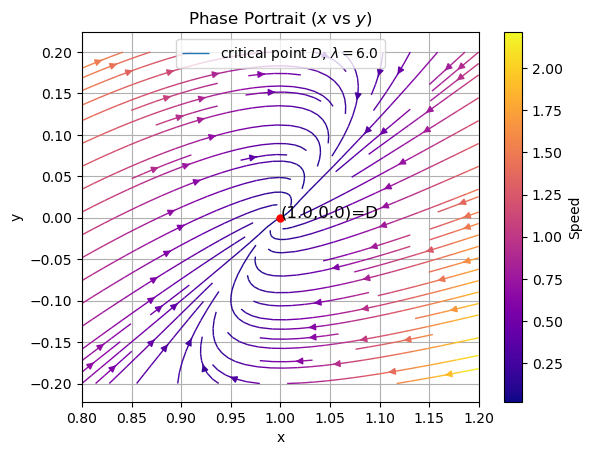}
	\caption{  Various plots for the critical point D. Left panel : case (f) with $\lambda=12$, middle panel: case (g) with $\lambda=10$ and right panel: case (i) with $\lambda =6$}
	\label{image_D}
	
\end{figure}
\begin{figure}
	\includegraphics[scale=0.5]{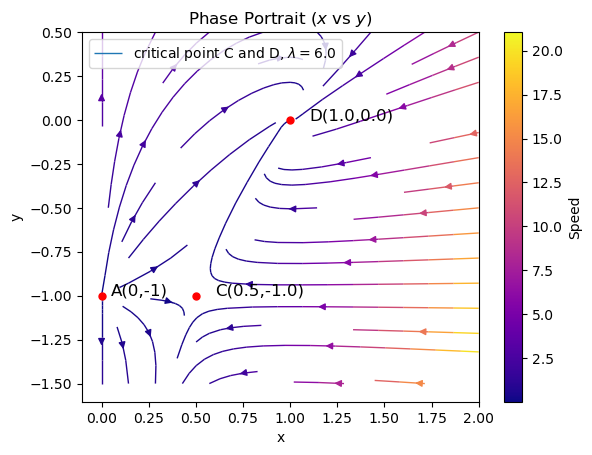}
		\includegraphics[scale=0.5]{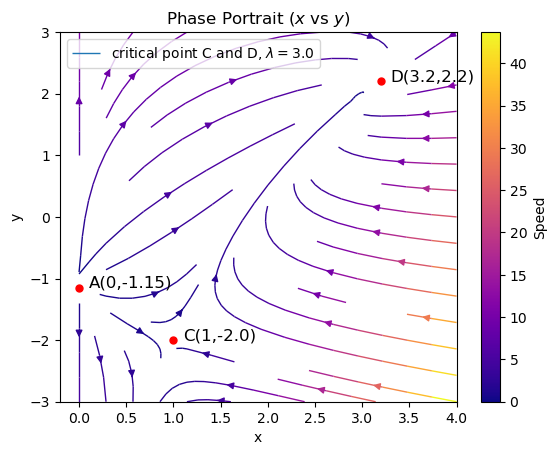}
	 
	\caption{Figure for evolution of x vs y for the critical points C and D with $\lambda = 6$ and $\lambda=3$.}  
	\label{image_CD}
\end{figure}

\textbf{Critical point D:}
The point ($\frac{12}{\lambda }-1, \frac{12}{\lambda }-2,0$) is the citical point for D with  the eigenvalues of the Jacobian matrix as $\left\{\frac{\lambda -12}{2},\lambda -10,\lambda -9\right\}$. Clearly four conditions can arise here : (f) for $\lambda  \ge 12$ with all three +ve eigenvalues and {$\Omega_\phi<0, \Omega_m>1$} (g) for $10\le \lambda<12 $ one -ve and two +ve eigenvalues (h) for $10<\lambda \le 9$ two -ve and one +ve eigenvalues  and {$\Omega_\phi<0, \Omega_m>1$},   (i)for  $\lambda < 9$ all three  eigen values are real and -ve.  All these points (f),(g) and (h) are having atleast one +ve eigenvalue. Interestingly, for $\lambda = 9$ the points C and D overlap.  $\lambda < 9$ give the eigenvalues of the critical points which are  both -ve.  $\lambda = 6$ gives us $\Omega_\phi=-1, \Omega_m = 0$. In the region $6<\lambda<9$ although the system is stable but from cosmological point of view this region is undesirable as   {$\Omega_\phi<0, \Omega_m>1$}. At $\lambda = 7.2 $ the matter domination is there as $\Omega_r=0 ,\Omega_\phi =0, \Omega_m = 1$. There is a brief window for $\lambda$ when it can give physically viable results. For $6<\lambda<7.2$ we get {$0<\Omega_\phi,\Omega_m<1$}. However, $\lambda<6$ gives the unphysical sector for all the energy densities.  So it gives a scalar field dominated universe at the beginning with no radiation or matter. Fig(\ref{image_CD}) shows such a point where the point D acts as an attractor point on the $z=0$ plane.  
\subsection{Evoluion of the density parametrs and the equation of state parameter:}
\begin{figure}
	\includegraphics[scale=0.35]{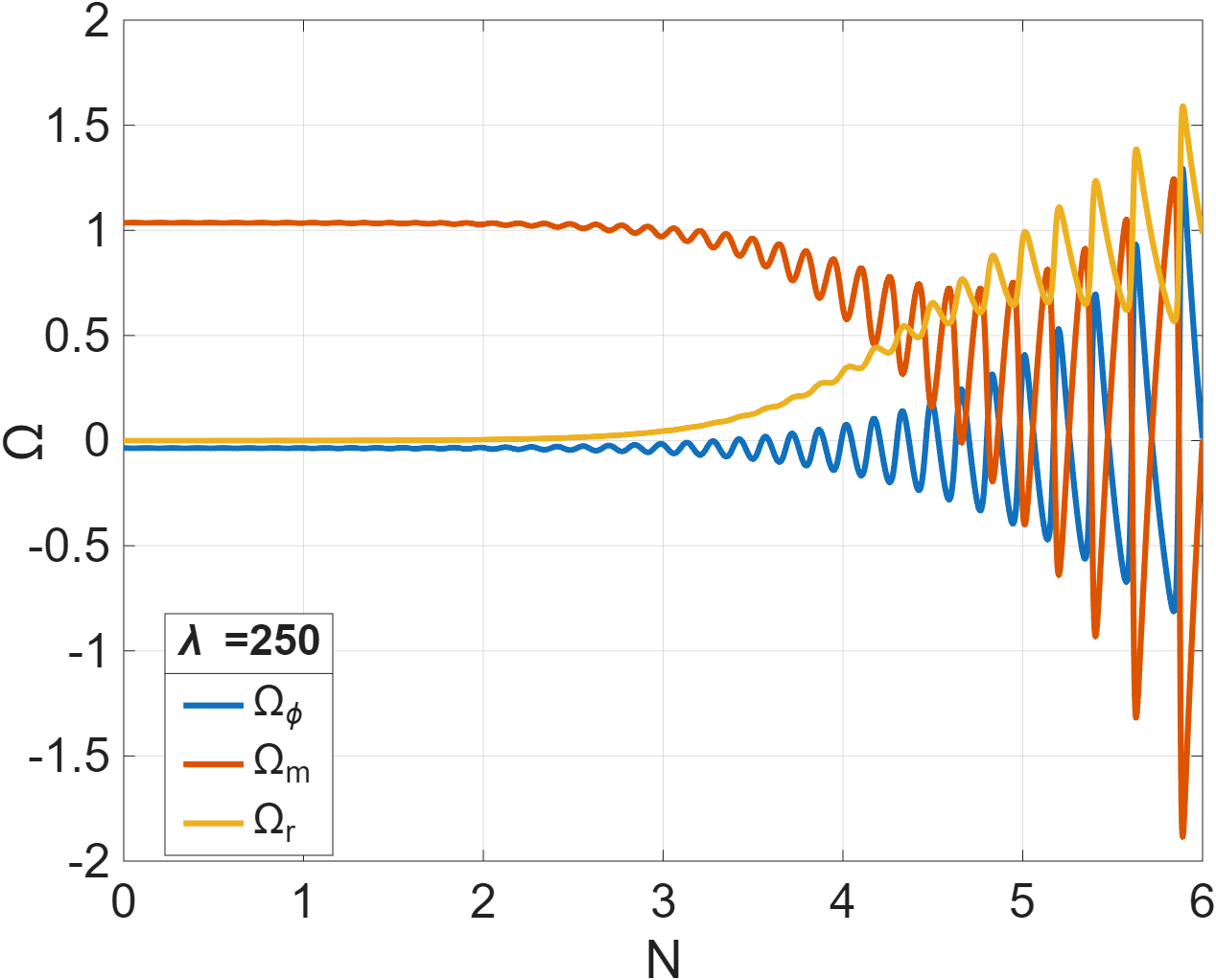}
	\includegraphics[scale=0.35]{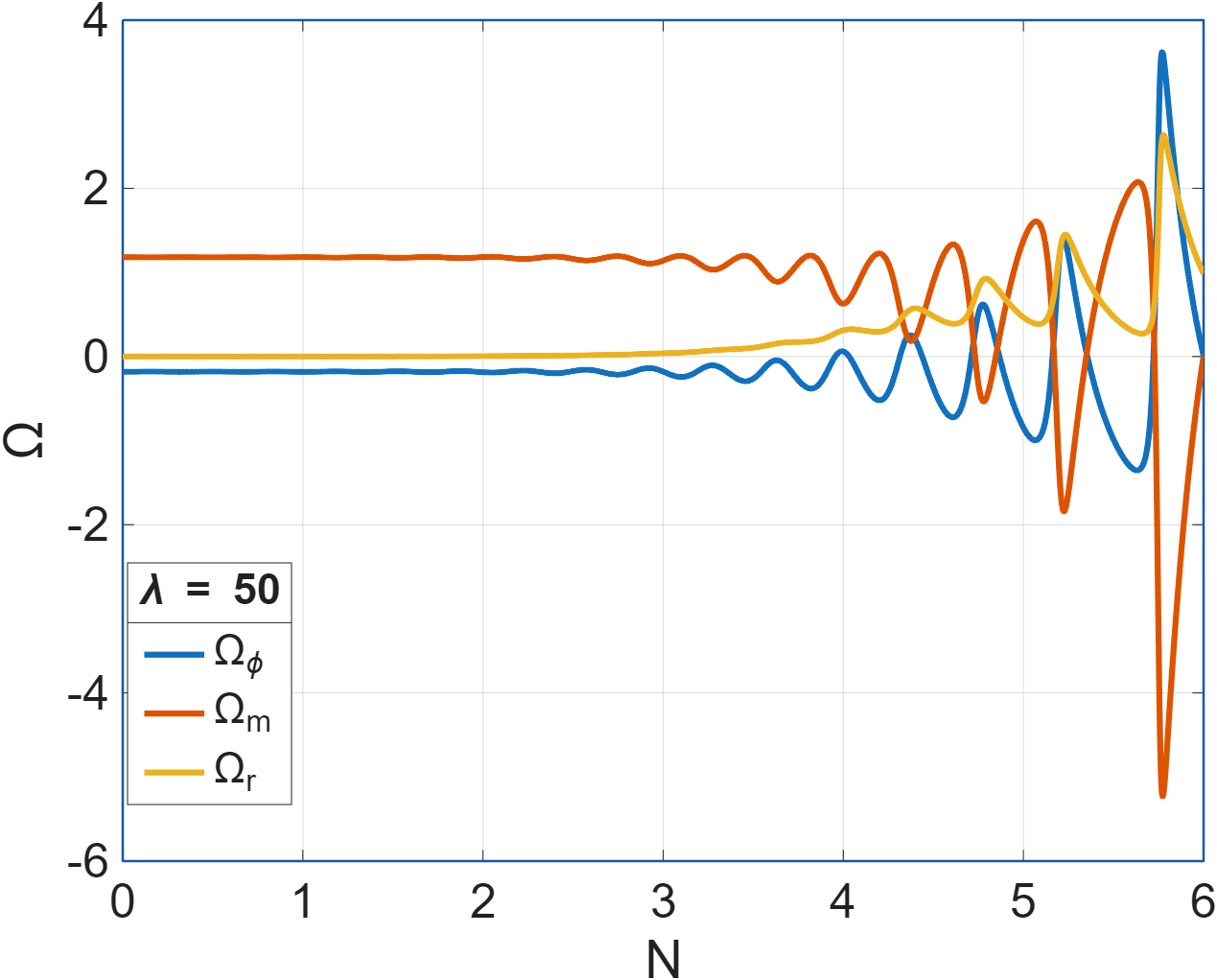}
	\caption{Variation of the relative energy parameters of various components (matter, radiation and the Cuscuton scalar field) . Left panel for $\lambda = 250$ and right panel for $\lambda = 50$ with the initial conditions  $x, y, z = [0.11e-6; 0.005 ; 0.985]$}
	\label{OmegaPlot}
\end{figure}

  In Fig (\ref{OmegaPlot}) we plot the density parameters $\Omega_\phi, \Omega_m$ and $\Omega_r$ for specifically two values of $\lambda$ by solving the dynamical equations (\ref{dynamical1} - \ref{dynamical3}). It is found that the radiation density parameter $\Omega_r$ dominates at the beginning and the oscillates to almost zero for both the $\lambda$ values. With increasing $N$, radiation density parameter becomes asymptotically zero faster. Note that, if we decrease the values of $\lambda$, specifically below 50, would give almost flat values of $\Omega_r$ and hence $\Omega_\phi$. This means the oscillatory nature will be lesser.  As an example, for $\lambda = 8$ there is a sudden increase of $\Omega_\phi >>1$ and to balance it $\Omega_m$ becomes much lesser than 1. This is also indicative from the two plots in Fig [\ref{OmegaPlot}].  

\begin{figure}
	\includegraphics[scale=.80]{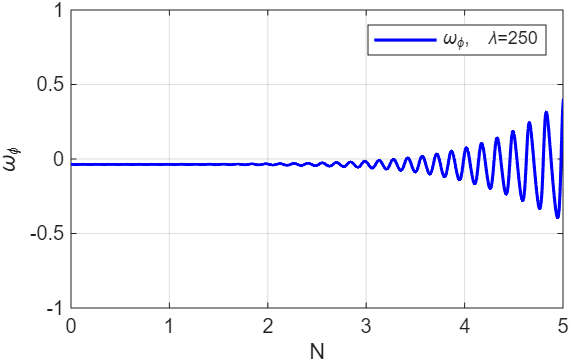}
	\includegraphics[scale= 0.8]{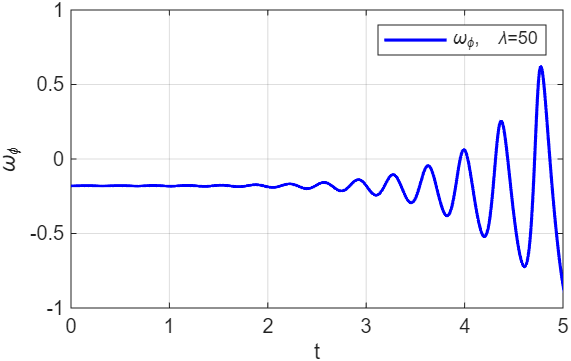}
	\caption{Variation of the equation of state parameter for the cuscuton field with $N = \ln a$. Left papel for $\lambda = 250$ and the right panel is for $\lambda  =50$ with initial conditions $x, y, z = [0.11e-6; 0.005 ; 0.985]$.}
	\label{omegaphiPlot}
\end{figure}

In Fig (\ref{omegaphiPlot}) we analyze the behavior of equation of state parameter with e-folding N.   The EoS parameter for $\phi$ field also shows a clear damping oscillation. $\omega_\phi$ oscillates above the phantom boundary $\omega_\phi = -1$. However, it is seen that, if we increase the $\lambda$ values the oscillation frequency increases and the damping is very faster. Smaller $\lambda$ values, typically lesser than 10, show much lesser values from the phantom crossing. Thus higher redshift  values are indicative of non-$\Lambda$CDM behavior in the early as well as in the late time.  

\section{Conclusion}\label{conclusion}
 since the inception of the theory \cite{Afshordi2007}, Cuscuton gravity has been an active field of research for quite a time  with numerous studies \cite{Andrade2019, Mylova2024, Bazeia2023, Bazeia2025, Iyonaga2018, Maeda2022}. Given its importance, it is crucial to explore these models from various perspectives. In this paper, we have focused on  the symmetry aspect, especially the implications of infinitesimal point transformations of the Cuscuton type theory. Since the point transformations are connected to borader classes of symmetry transformations,  like the Lie-B\.acklund transformation, identifying the corresponding symmetry parameters is essential. To achieve this, we have first identified the true degrees of freedom by analyzing the higher derivative nature of the theory. Using the first order formalism, we have analyzed the corresponding constraint structure of the model and found that a total all together 8 second class constraints exist there. The true degrees of freedom are found out to be \textit{two} consistent  with other related models of cucuton \cite{Gomes2017, Mukohyama2019}.  The reduction is possible because we can identify the surface term in the form of total derivative.  We now perform  the infinitesimal point transformation of the coordinates and fields to construct the corresponding generator. According to Noether's second theorem, the first prolongation of the Lagrangian and the conserved charge is related due to the variational principle \cite{BlumanB, Christodoulakis2014, Capozziello1997}. Using this approach, we have found out the corresponding infinitesimal symmetry parameters along with the form of the potential.  Interestingly, the  consistency equations also demand that the coefficient of the original Cuscuton should vanish $a_2 = 0$ ;  otherwise, the model do not agree with Noether's second theorem. We also have explored the Killing symmetries of this model. We have found out the Killing vectors and Killing tensors corresponding to the metric formed due to the kinetic matrix  after removing the surface term and imposing the condition $a_2 = 0$. We have computed the associated charges and verify their conservation through time preservation, confirming the presence of conserved quantities in the system.
 
Next, we have explored the cosmology of the model.  Starting from the equivalent Friedmann equations, we have identified the dimensionless constraints to perform dynamical analysis.The resulting dynamical equations reveal four fixed points. Out of these four fixed points, two are always unstable while two are conditionally stable. The  parameter $\lambda$ plays an important role in determining the stability properties of the model. We also have calculated numerically the evolution of the energy densities corresponding to radiation, matter and for the field $\phi$. The matter density was prevailing in the past which is clearly in contrary to the present observable Universe. It is imperative from the plots that they  show damped oscillatory behavior of the energy density parameters with time. Similarly, the evolution of equation of state parameter also shows oscillatory behavior indicating a cyclic Universe undergoing contractions and expansions within the phantom range.  Such  cyclic behavior of the Universe  have already been studied   \cite{Steinhardt2002, Chang2012, Steinhardt_2002a}. However, the present model  deviates from the interpretation by the authors  \cite{Panpanich2023}  due to consideration of Lie symmetry for which we have imposed the condition $a_2 = 0$. This condition fundamentally alters the model?s dynamics and cosmological implications.
 
 It is agreeable that the present model despite showing unwarranted cosmological behavior for the energy densities is interesting for showing how the behavior changes once we incorporate the symmetry point of view - specifically the infinitesimal Lie symmetries. This perspective opens up intriguing possibilities for further research, encouraging the exploration of other variations of the C uscuton model under the framework of Lie symmetry analysis. 
 %\bibliographystyle{siam}
%\bibliography{bibliographies.bib}

\end{document}